\documentclass{article}
\usepackage{natbib,epsfig,multirow}

\usepackage{amsthm,amsmath,amssymb,colonequals,multirow,graphicx,epstopdf}

\newcommand{\tr}{\,\mbox{tr}}
\newcommand{\load}{\mathbf\Lambda}
\newcommand{\noisev}{\mathbf\Psi}

\newcommand{\ident}{\mathbf{I}}
\newcommand{\vecY}{\mathbf{Y}}
\newcommand{\vecy}{\mathbf{y}}

\newcommand{\vecx}{\mathbf{x}}

\newcommand{\matzero}{\mathbf{0}}

\newcommand{\mata}{\mathbf{A}}
\newcommand{\vecxc}{\mathbf{X}}
\newcommand{\vecX}{\mathbf{X}}

\newcommand{\vecuc}{\mathbf{U}}
\newcommand{\vecz}{\mathbf{z}}

\newcommand{\vecZ}{\mathbf{Z}}
\newcommand{\vecS}{\mathbf{S}}
\newcommand{\thet}{\mathbf{\Theta}}

\newcommand{\vectheta}{\mbox{\boldmath$\theta$}}

\newcommand{\varthet}{\mbox{\boldmath$\vartheta$}}
\newcommand{\vecmu}{\mbox{\boldmath$\mu$}}
\newcommand{\vecbeta}{\mbox{\boldmath$\beta$}}

\newcommand{\matsig}{\mbox{\boldmath$\Sigma$}}
\newcommand{\mSigma}{\mbox{\boldmath$\Sigma$}}
\newcommand{\mTheta}{\mbox{\boldmath$\Theta$}}
\newcommand{\mXi}{\mbox{\boldmath$\Xi$}}
\newcommand{\vecxi}{\mbox{\boldmath$\xi$}}
\newcommand{\vecsigma}{\mbox{\boldmath$\sigma$}}

\newcommand{\mLambda}{\mbox{\boldmath$\Lambda$}}
\newcommand{\mPsi}{\mbox{\boldmath$\Psi$}}

\date{}
\begin{document}

\title{A Partial EM Algorithm for Clustering White Breads}
\author{Ryan P.\ Browne\thanks{Department of Mathematics \& Statistics, University of Guelph, Guelph, Ontario, N1G 2W1, Canada. E-mail: \{rbrowne,paul.mcnicholas\}@uoguelph.ca.} , Paul D.\ McNicholas$^*$, Christopher J.\ Findlay\thanks{Compusense Inc., Guelph, Ontario, N1G 4S2, Canada. E-mail: cfindlay@compusense.com}}
\hyphenation{monotonicity}
\maketitle

\begin{abstract}
The design of new products for consumer markets has undergone a major transformation over the last 50 years. Traditionally, inventors would create a new product that they thought might address a perceived need of consumers. Such products tended to be developed to meet the inventors own perception and not necessarily that of consumers. The social consequence of a top-down approach to product development has been a large failure rate in new product introduction. By surveying potential customers, a refined target is created that guides developers and reduces the failure rate. Today, however, the proliferation of products and the emergence of consumer choice has resulted in the identification of segments within the market. Understanding your target market typically involves conducting a product category assessment, where 12 to 30 commercial products are tested with consumers to create a preference map. Every consumer gets to test every product in a complete-block design; however, many classes of products do not lend themselves to such approaches because only a few samples can be evaluated before `fatigue' sets in. We consider an analysis of incomplete balanced-incomplete-block data on 12 different types of white bread. A latent Gaussian mixture model is used for this analysis, with a partial expectation-maximization (PEM) algorithm developed for parameter estimation. This PEM algorithm circumvents the need for a traditional E-step, by performing a partial E-step that reduces the Kullback-Leibler divergence between the conditional distribution of the missing data and the distribution of the missing data given the observed data. The results of the white bread analysis are discussed and some mathematical details are given in an appendix.\\

\noindent\textbf{Keywords:} Balanced-incomplete-block; mixture models; progressive EM; sensometrics; white bread.
\end{abstract}

\section{Introduction}\label{sec:intro}

Consumer-driven product development of new consumer products and the improvement of existing products have become recognized as a best-practice in industry. Food researchers have become increasingly dependent on understanding consumer wants and desires to effectively design food products \citep{jaeger03}. To understand consumer behaviour, preference maps are built by assessing consumer liking of an appropriate range of commercial products within a category. From these liking data, a model may be built that describes the ÒidealÓ product for the test population. However, most product categories will have more than a single ideal product, with two or more liking clusters revealed. Hedonic taste tasting is the most common practice used to measure consumer liking within a target population \citep{lawless10}. In a complete-block design, every consumer gets to taste every product, but many product categories do not facilitate this sampling plan. When tasting wine, for example, a consumer can only evaluate three or four samples before `fatigue' sets in, compromising the quality of the data collected. The fact that consumers tend to behave like experts puts into doubt the value of data obtained over multiple tasting sessions \citep{findlay08}. Therefore, a balanced-incomplete-block (BIB) design is used for high-fatigue products. The resulting data are sparse and tend to be heterogenous; therefore, we must identify sub-populations in an incomplete-data setting.


 
In this paper, we consider an analysis of 12 white breads. The descriptive analysis of the breads provides a measure of the range of sensory properties found within the product category. It also gives us information regarding changes that may improve the sensory liking of a product for a specific consumer segment. Consumer research is conducted to measure liking on a nine-point hedonic scale anchored at dislike extremely (1) and like extremely (9), with the midpoint (5) indicating neither like nor dislike. By clustering consumers on similarity of liking profiles across the products, it is possible to determine the sensory attributes that contribute to like/dislike within each cluster.
%
The calibrated descriptive analysis was performed by a trained panel using well-defined product attributes that have been rated for intensity on a scale from $0$ to $100$. The attributes are generated to provide a complete sensory description of the breads. Some attributes are found at low intensity, but are important in differentiating products. There are also attributes that are defined by a major attribute or group of attributes. For example, sourdough breads would score high in sourness and sour aroma and flavour.
%
The products selected for this study encompass the commercial sliced white bread category. The range goes from the extremely popular sandwich breads that are fine-celled, spongy, and bland to a ciabatta-style Italian hearth bread. The breads differ in crust colour and roughness, texture of the crumb, and flavour, but all fall within the realm of sliced white bread. A total of 369 consumers evaluated six breads within a 12-present-6 BIB design. 

One straightforward way to tackle such an incomplete-data problem is to impute the missing data prior to the analysis. However, this approach is not generally desirable for clustering problems because the imputed values will be partly based on data from other sub-populations. Herein, we develop a clustering approach for these data based on a finite Gaussian mixture model. A random variable $\vecxc$ follows a $G$-component finite Gaussian mixture model if its density can be written
\begin{equation} \label{eqn:gmm}
f(\vecx\mid\varthet)= \sum_{g=1}^G \pi_g\phi(\vecx\mid\vecmu_g,\matsig_g),
\end{equation} where $\pi_g>0$, with $\sum_{g=1}^G\pi_g=1$, are mixing proportions, $\phi(\vecx\mid\vecmu_g,\matsig_g)$ is multivariate Gaussian density with mean $\vecmu_g$ and covariance matrix $\matsig_g$, and $\varthet=(\pi_1,\ldots,\pi_G,\vecmu_1,\ldots,\vecmu_G,\matsig_1,\ldots,\matsig_G)$. Finite mixture models have been used for clustering for at least fifty years \citep{wolfe63} and such applications are commonly referred to as `model-based clustering' \citep[cf.][]{fraley02a}. One problem with applications of Guassian mixture models is the number of free covariance parameters: $Gp(p+1)/2$. To overcome this, many authors have considered imposing constraints on decomposed component covariance matrices \citep[e.g.,][]{celeux95} and other have considered underlying latent factor models \citep[e.g.,][]{ghahramani97}. We consider an underlying latent factor model herein (cf.\ Section~\ref{sec:meth}) and because so many of the data are missing, we assume common component covariance. The result is a parsimonious Gaussian mixture model.

The expectation-maximization (EM) algorithm \citep{dempster77} is the standard approach to parameter estimation for model-based clustering \citep[cf.][]{mclachlan88}. However, in our incomplete-block data we obtain only $6$ liking scores from $12$ products for each consumer; therefore, an EM algorithm would require ${12 \choose 6}=924$ different $6\times 6$ matrix inversions for each mixture component in each E-step.  To circumvent this problem, we develop a `partial' EM (PEM) algorithm that requires only a single $12\times 12$ matrix inversion for each mixture component. We show that this PEM retains the monotonicity property and thus all of the convergence properties of the standard EM algorithm, but is much more computationally efficient than the standard EM algorithm for this particular problem. 


The remainder of this paper is laid out as follows. 
In Section~\ref{sec:back}, we review the application of the EM algorithm for missing data. Then, our parsimonious Guassian mixture model is presented and the PEM algorithm is developed (Section~\ref{sec:meth}). We apply our method to the white bread data in Section~\ref{sec:white}, where we also compare our PEM algorithm to the standard EM algorithm (Section~\ref{sec:sim}). The paper concludes with discussion and suggestions for future work (Section~\ref{sec:disc}). 

%
%
%

\section{The EM Algorithm for Missing Data Problems}\label{sec:back}
 
The EM algorithm is an iterative procedure for finding maximum likelihood estimates when data are incomplete. Therefore, EM algorithms are naturally suited for missing data problems. The EM algorithm consists of alternating between E- and M-steps until a convergence criterion is satisfied. In the E-step, the expected value of the complete-data (i.e., the observed plus missing data) is computed, and in the M-step, this quantity is maximized with respect to the parameters. Formally, the EM algorithm is a special case of an MM algorithm of the minorization-maximization variety \citep{hunter00,hunter04}.  

Suppose we observe $p$-dimensional $\vecy_1,\ldots,\vecy_n$ such that each $ \vecy_i$ can be decomposed into an observed component, $\vecx _i$, of dimension $m$, a missing component, $\vecz _i$, of dimension $l=p-m$, and 
\begin{equation*}
 \vecY_i = 
 \left[\begin{array}{c}  \vecX_i \\ \vecZ_i  \end{array}\right] 
 \backsim
 \mathcal{N} \left( \vecmu =
 \left[\begin{array}{c} \vecmu _x \\ \vecmu_z \end{array}\right] , 
\mSigma = \left[ \begin{array}{cc} \mSigma_{xx} & \mSigma_{xz} \\ \mSigma_{xz} & \mSigma_{zz}\end{array}\right] 
 \right).
\end{equation*}
Then, the conditional distribution of the missing data given the observed is given by
\begin{equation*}
\vecZ_i | \vecX_i = \vecx_i \backsim \mathcal{N}(\vecmu_{z_i.x_i} \colonequals \vecmu_z + \mSigma_{zx}\mSigma_{xx}^{-1} \left(\vecx_i - \vecmu_x \right), \mSigma_{z.x} \colonequals \mSigma_{zz} - \mSigma_{zx}\mSigma_{xx}^{-1} \mSigma_{xz}).
\end{equation*}
Now, define
\begin{equation} \label{conditional mean}
\hat{\vecz}_i := \vecmu_{z.x_i}= \vecmu_z + \mSigma_{zx}  \mSigma_{xx}^{-1} \left(  \vecx_i - \vecmu_x \right),
\end{equation}
\begin{equation} \label{conditional covariance}
\hat{\vecY}_i :=  
\left[\begin{array}{cc} \matzero_{l \times l} & \matzero_{l \times m} \\ \matzero_{l \times m} & \hat{\vecZ}_i \end{array}\right]
= \left[\begin{array}{cc} \matzero_{l \times l} & \matzero_{l \times m} \\ \matzero_{m \times l} & \mSigma_{z.x}  \end{array}\right]
= \left[\begin{array}{cc} \matzero_{l \times l } & \matzero_{l \times m} \\ \matzero_{m \times l} & \mSigma_{zz} - \mSigma_{zx} \mSigma_{xx}^{-1} \mSigma_{xz}   \end{array}\right],
\end{equation}
where $ \matzero_{l \times m}$ is an $l \times m$ matrix of zeros and $\hat{\vecy}_i = ( \vecx_i, \hat{\vecz}_i )$. Using this notation, the EM updates for the mean and covariance can be written as 
\begin{equation} \label{mean and covariance updates}
\hat{\vecmu}^{(t+1)} = \overline{ \vecy } =\frac{1}{n} \sum_{i=1}^n  \hat{\vecy}_i 
\;\;\;\;\;\;\;
\mbox{and}
\;\;\;\;\;\;\;
\hat{\mSigma}^{(t+1)} = \vecS = \sum_{i=1}^n \left( \hat{\vecy}_i - \overline{ \vecy } \right) \left(  \hat{\vecy}_i -   \overline{ \vecy } \right)'   + \sum_{i=1}^n \hat{\vecY}_i,
\end{equation}
respectively.
Because our white bread data have lots of missing observations, `standard' E-steps are very computationally expensive. For example, each observation requires inversion of a $6\times 6$ matrix; this amounts to ${12 \choose 6} = 920$ different matrix inversions at each iteration. 

Next, consider approximate E-steps instead of full E-steps. 
All of these procedures will work with the inverse of $\mSigma$ and its principal sub-matrices and vectors. We will assume that $\mSigma^{-1}$ is known; this quantity is typically readily available because it is necessary to calculate the log-likelihood at each EM iteration. We denote 
\begin{equation*}
\mSigma = \left[\begin{array}{cc} \mSigma_{xx} & \mSigma_{xz} \\ \mSigma_{xz} & \mSigma_{zz}\end{array}\right]
\;\;\;\;\;\;\;\;\;\;
\mbox{and}
\;\;\;\;\;\;\;\;\;\;
\mSigma^{-1} = \mXi = \left[\begin{array}{cc} \mXi_{xx} & \mXi_{xz} \\ \mXi_{zx} & \mXi_{zz}\end{array}\right]
\end{equation*}
because there exists a relationship between $ \mXi_{zz}$ and $\mSigma_{z.x}$. 
This relationship exists because $\mSigma_{z.x}$ is the Suhur complement of the matrix $\mSigma$ which has the property that 
\begin{equation*}
  \mSigma_{z.x} = \mXi_{zz}^{-1} 
\;\;\;\;
\mbox{ and, equivalently,  }
\;\;\;\;
  \mSigma_{z.x}^{-1} = \mXi_{zz} .
\end{equation*}
In addition, there exists a relationship between the regression coefficients and $\mXi$, which can be derived through block inversion of $\mSigma$,
\begin{equation*}
 \mSigma_{zx}  \mSigma_{xx}^{-1} = - \mXi_{zz}^{-1} \mXi_{zx}.
\end{equation*}
These relations can be useful if the dimension of $\vecz$ is smaller than the dimension of $\vecx$. For example, if there is a single missing observation then using $\mSigma$ requires a $p-1$ matrix inversion, whereas if $\mXi$ is known we only require an inversion of a $1 \times 1$ matrix.

For the extension $G>1$ mixture components, we require the weighted versions of (\ref{mean and covariance updates}). The weight for observation $i$ in component~$g$ is 
\begin{equation}
w_{ig} = \frac{\pi_g\phi\left(\vecx_i | \vecmu_{g, x}, \mSigma_{g,xx}\right)}{\sum_{k=1}^G \pi_k \phi\left(\vecx_i | \vecmu_{k, x}, \mSigma_{k,xx} \right)},
\end{equation}
where $\hat{\pi}_g^{(t+1)} = {n_g}/{n} = ({1}/{n})\sum_{i=1}^n w_{ig}$,
$\hat{\vecmu}_g^{(t+1)} = \overline{ \vecy }_g =({1}/{n_g})\sum_{i=1}^nw_{ig}\hat{\vecy}_i$, and 
\begin{equation} \label{group mean and covariance updates}
\hat{\mSigma}^{(t+1)}_g = \vecS_g = \frac{1}{n_g} \sum_{i=1}^n \left[ w_{ig} \left( \hat{\vecy}_i - \overline{ \vecy } \right) \left(  \hat{\vecy}_i -   \overline{ \vecy } \right)'  +  \widehat{\vecY}_i \right],
\end{equation}
where $n_g = \sum_{i=1}^n w_{ig}$. 

\section{Methodology}\label{sec:meth}

\subsection{Finite Mixture Models with Common Factors}

As already mentioned (Section~\ref{sec:intro}), it is common practice to introduce parsimony via constrained component covariance matrices. When dealing with sparse data, such as the white bread data, estimating the covariance matrix can be especially difficult. 
Therefore, we use a variant of the mixture of factor analyzers model \citep{ghahramani97,mclachlan00a} in which we constrain the component factor loading matrices to be equal across groups \citep[cf.][]{mcnicholas08}. The factor analysis model \citep{spearman04,bartlett53} assumes that a $p$-dimensional random vector~$\vecxc_i$ can be modelled using a $q$-dimensional vector of latent factors $\vecuc_i$, where $q\ll p$. The model can be written $\vecxc_i = \vecmu + \load\vecuc_i + \mbox{\boldmath$\epsilon$}$, where $\load$ is a $p\times q$ matrix of factor weights, the latent variables $\vecuc_i \sim \mathcal{N}(\mathbf{0},\ident_q)$, and $\mbox{\boldmath$\epsilon$} \sim \mathcal{N}(\mathbf{0},\noisev)$, where $\noisev$ is a $p \times p$ diagonal matrix. Therefore, the marginal distribution of $\vecxc_i$ is $\mathcal{N}(\vecmu,\load\load'+\noisev)$. It follows that the density for the mixture of factor analyzers model is that of Equation~\ref{eqn:gmm} with $\matsig_g = \mLambda_g\mLambda_g' + \mPsi_g$. The model we use for the analysis of the bread data assumes equal factor loading matrices across components, i.e., $\matsig_g = \mLambda\mLambda' + \mPsi_g$, and so the EM algorithm updates are given by
%
%
\begin{equation}
\mbox{vec}\left(  \hat{\mLambda}^{(\text{new})} \right) = 
\left[  \sum_{g=1}^G n_g \hat{\mPsi}_g^{-1} \otimes  \hat{\thet}_g  \right]^{-1}
 \mbox{vec} \left( \sum_{g=1}^G n_g \hat{\mPsi}_g^{-1} \vecS_g \vecbeta_g'   \right),
\end{equation}
\begin{equation}
\hat{\mPsi}^{(\text{new})}_g= \mbox{diag}\left\{ \mathbf{S}_g - 2\hat{\mLambda}^{(\text{new})} \hat{\vecbeta}_g \mathbf{S}_g  +  \hat{\mLambda}^{(\text{new})} \thet_g \left( \hat{\mLambda}^{(\text{new})}\right)' \right\}.
\end{equation}

\subsection{PEM Algorithm}

We follow \cite{neal1998} and store the sufficient statistics $(\widehat{\vecz}_1, \ldots, \widehat{\vecz}_n)$ and $(\widehat{\vecZ}_1, \ldots, \widehat{\vecZ}_n)$. However, instead of doing a complete update of a partial set of the sufficient statistics as suggested by \cite{neal1998}, we perform a partial E-step at each iteration. These partial E-steps can be shown to reduce the Kullback-Leibler (KL) divergence at every step, which ensures that the monotonicity of the EM algorithm is preserved. 
From \cite{neal1998}, the EM algorithm can be viewed as minimizing the function 
\begin{equation*}
\sum_{i=1}^n F( \widetilde{P}_i, \theta) = - \sum_{i=1}^n D( \widetilde{P}_i ||  P_{i,\theta} ) + \sum_{i=1}^n l( \vecx_i | \theta),
\end{equation*}
where $D( \widetilde{P}_i ||  P_{i,\theta} )$ is the KL divergence between the distribution of the missing data $\widetilde{P}_i$ and the conditioal distribution of the missing data given the observed data,  $P_{\theta} =P(\vecZ_i | \vecx_i, \theta)$. A `standard' E-step sets $\widetilde{P}_i$ to $P(\vecZ_i | \vecx_i, \theta_t)$, for all $i$, at each iteration $t$. \cite{neal1998} suggest a partial or sparse E-step, where a subset of $\widetilde{P}_i$ is updated to $P(\vecZ_i | \vecx_i, \theta_t)$ at each EM iteration. The algorithm we describe in the next two sections partially updates each $\widetilde{P}_i$ towards $P(\vecZ_i | \vecx_i, \theta)$ such that the KL divergence is reduced but not minimized. 
For the multivariate Gaussian distribution and a particular $i$, the EM algorithm can be viewed as minimizing  
\begin{equation*}
 F( N_\vecz, \vecx_i, \theta ) = - D_{\mbox{KL}} \left( N_\vecz || N_{\vecz.\vecx_i} \right)+ l(\vecx_i | \theta),
\end{equation*}
with respect to $N_z$, the distribution of the latent or missing variables, and the parameter set $\theta$.


The KL divergence between the missing data distribution with mean $\hat{\vecz}_i$ and variance $\widehat{\vecZ}_{i}$ and the conditional distribution of the missing data given the observed data $\vecx_i$  is 
\begin{equation} \label{KL of missing data} 
D_{\mbox{KL}} \left(N_z || N_{z.x_i} \right) = \frac{1}{2} \left[\mbox{tr}\left\{\mSigma_{z.x}^{-1} \widehat{\vecZ}_{i}\right\}  + (\hat{\vecz}_i-\vecmu_{z.x})'\mSigma_{z.x}^{-1}(\hat{\vecz}_i-\vecmu_{z.x}) - \ln \left(\frac{| \widehat{\vecZ}_{i}|}{|\mSigma_{z.x}|} \right)\right].
\end{equation}
From Equation~\ref{KL of missing data}, we can see that if we set $\hat{\vecz}_i$ to the conditional mean and $\widehat{\vecZ}_{i}$ to the conditional covariance matrix, then the KL divergence is minimized. However, for our data this involves inverting ${12 \choose 6}=924$ different $6\times 6$ matrices at each iteration. Finding the minimum distribution for the missing data in each row is computationally expensive, so we will instead iteratively minimize the KL divergence on simpler computations. 

\subsection{Notation}
Hereafter, the following notation will be used. Let $\mSigma_j$ be the principal sub-matrix of $\mSigma$, obtained by deleting column $j$ and row $j$. Let $\sigma_{j}$ and $\xi_{j}$ be the $j$th diagonal elements of $\mSigma$ and $\mXi$, respectively. Let $\vecsigma_{j}$ and $\vecxi_{j}$ be the $j$th rows of $\mSigma$ and $\mXi$, respectively, with the $j$th element deleted. For example, if $j=1$, then
\begin{equation} \label{principal sub-matrix}
\mSigma = \left[\begin{array}{cc} \sigma_{1} & \vecsigma_{1}'  \\ \vecsigma_{1} & \mSigma_{1}\end{array}\right]
\;\;\;\;\;\;\;\;\;\;
\mbox{and}
\;\;\;\;\;\;\;\;\;\;
\mSigma^{-1} = \mXi = \left[\begin{array}{cc} \xi_{1} & \vecxi_{1}'  \\ \vecxi_{1} & \mXi_{1}\end{array}\right].
\end{equation}

\subsection{Minimizing the KL Divergence With Respect to  $\widehat{\vecz}_i$}

To minimize the KL divergence with respect to $\widehat{\vecz}_i$, we set it to $\vecmu_{z.x}$, which is given in (\ref{conditional mean}). However, $\vecmu_{z.x}$ depends on a matrix inversion that depends on $i$. We now develop an updating equation that  reduces the number of matrix inversions.
The KL divergence depends on $\widehat{\vecz}_i$ through one term in equation (\ref{KL of missing data}) and 
\begin{equation*}
\underset{\hat{\vecz}_i}{\operatorname{argmax}} \;
D_{\mbox{KL}} \left( N_z || N_{z.x_i} \right)
=
\underset{\hat{\vecz}_i}{\operatorname{argmax}} \; \left( \hat{\vecy}_i - \vecmu\right)' \mSigma^{-1} \left( \hat{\vecy}_i - \vecmu\right) 
\end{equation*}
because the last term in 
\begin{equation} \label{quadratic vector result}
 \left( \hat{\vecy}_i - \vecmu\right)' \mSigma^{-1} \left( \hat{\vecy}_i - \vecmu\right) 
 = (\hat{\vecz}_i-\vecmu_{z.x})'\mSigma_{z.x}^{-1}(\hat{\vecz}_i-\vecmu_{z.x})
  + ( \vecx_i-\vecmu_{x})'\mSigma_{xx}^{-1}(\vecx_i-\vecmu_{x})  
\end{equation}
is fixed and, moreover, does not depend on $\vecz_i$ (cf.\ Appendix \ref{vector quadratic form}).

We now consider a conditional minimization, by each element in $\vecy_i= (y_{i1}, \ldots, y_{in})$, of 
\begin{equation*} 
\underset{\hat{\vecz}_i}{\operatorname{argmax}} \; \left( \hat{\vecy}_i - \vecmu\right)' \mSigma^{-1} \left( \hat{\vecy}_i - \vecmu\right), 
\end{equation*}
which is given by
\begin{equation*}
y_{ij}^{(t)} = \left\{ \begin{array}{cc} 
\mu_j + \vecsigma_{j} \mSigma_{j}^{-1} \left( y_{-j}^{(t)} -\vecmu_{-k}  \right) & \mbox{if $j$ is corresponds to an element in } \widehat{\vecz}_i , \\ 
y_{ij} & \mbox{if $j$ is corresponds to an element in } \vecx_i , \end{array} \right.
\end{equation*}
where $\mu_k$ is the $k$th element of $\vecmu$. This procedure requires inversion of the $\mSigma_j$, i.e., the $j$th principal sub-matrix; which involves a $p-1$ matrix inversion. However, under the assumption that $\mSigma^{-1} = \mXi$ is known, the updates can be simplified to 
\begin{equation*}
y_{ij}^{(t)} = \left\{ \begin{array}{cc} 
\mu_j - \frac{1}{\xi_j} \vecxi_{j} \left( y_{-j}^{(t)} -\vecmu_{-j}  \right) & \mbox{if $j$ is corresponds to an element in } \widehat{\vecz}_i , \\ 
y_{ij} & \mbox{if $j$ is corresponds to an element in } \vecx_i. \end{array} \right.
\end{equation*}
This set of updates requires the inversion of a $1\times1$ matrix, which is trivial. These updates are guaranteed to converge to the global minimum because the objective function is convex. 
The advantage of this conditional minimization of the KL divergence is that the update is computationally simple. If we let the $n \times p$ matrix $\hat{\mata} = \left(\hat{\vecy}_1', \ldots, \hat{\vecy}_n'\right)$, then our partial E-steps update $\hat{\mata}$  by column instead of `standard' E-steps that update $\hat{\mata}$ by row.  

\subsection{Minimizing the KL Divergence With Respect to  $\widehat{\vecZ}_i$ }

Now, minimizing the KL divergence with respect to  $\widehat{\vecZ}_i$ we have
\begin{equation*}
\underset{\hat{\vecZ}_i}{\operatorname{argmax}} \;
D_{\mbox{KL}} \left( N_z || N_{z.x_i} \right) 
=
\underset{\hat{\vecZ}_i}{\operatorname{argmax}} \;
\tr\left\{\mSigma_{z.x}^{-1} \widehat{\vecZ}_{i} \right\} - \ln \left( \frac{ | \widehat{\vecZ}_{i}| }{ | \mSigma_{z.x} | }\right).
\end{equation*}
The log-determinant and the trace function are convex with respect to the positive definite matrices \citep{magnus1988}. Now, consider the function
\begin{equation}
\gamma\left(\vecZ_i \right)
 = 
\tr \left[  \left( \mSigma - \widehat{\vecY}_{i} \right) \mSigma^{-1}  \left( \mSigma - \widehat{\vecY}_{i} \right)\right],
\end{equation}
which is also convex, and these functions have the property
\begin{equation}
\underset{\hat{\vecZ}_{i}}{\operatorname{argmax}} \;
D_{\mbox{KL}} \left( N_z || N_{z.x_i} \right) 
=
\underset{\hat{\vecZ}_{i}}{\operatorname{argmax}} \;
\tr \left[  \left( \mSigma - \widehat{\vecY}_{i} \right) \mSigma^{-1}  \left( \mSigma - \widehat{\vecY}_{i} \right)\right]. 
\end{equation}
Both objective functions are minimized by the Schur complements (c.f.\ Appendix~\ref{matrix quadratic form} for the function $\gamma$). In addition, if one objective function is reduced then so is the other because both functions are convex and have the same (global) minimum. Therefore, if the function $\gamma$ is reduced at every iteration, then the KL divergence is reduced at every iteration and so the algorithm has the monotonicity property.

Conditional updates for the function $\gamma$ are derived in Appendix~\ref{matrix quadratic form}. We update $\hat{\vecY}_{i}$ by column and row, while holding the associated principal sub-matrix fixed. Therefore, if we denote the $j$th row of $\hat{\vecY}_i$ by $\hat{\vecY}_{i,j}$ and the $(p-1) \times p$ matrix obtained from removing the $j$th row as $\hat{\vecY}_{i,-j}$, then the updates can be written as 
\begin{equation*}
\hat{\vecY}_{i,j}^{(t+1)} = \left\{ \begin{array}{cc} 
\vecsigma_j + \left( \vecsigma_j - \hat{\vecY}_{i,j}^{(t)}  \right)' \mSigma_{j}^{-1} \left( \mSigma_{-j} - \hat{\vecY}_{i,-j}^{(t)}  \right)  & \mbox{if $j$ is associated with } \hat{\vecz}_i , \\ 
\matzero_{1\times p}& \mbox{if $j$ is associated with } \vecx_i , \end{array} \right.
\end{equation*}
and then we set the $j$th column of $\widehat{\vecY}_{i}^{(t+1)}$ equal to the $j$th row.

We can avoid the matrix inversion of the principal sub-matrix $\mSigma_{j}$ by again exploiting  the properties of the inverse of $\mXi$. Specifically, if $\vecxi_i' \vecsigma_i \neq 1$, then
\begin{equation*}
\mSigma_j^{-1} = \left[ \ident_{p-1} + \frac{1}{1-\vecxi_j' \vecsigma_j } \vecxi_j \vecsigma_j' \right] \mXi_j,
\end{equation*}
where $\ident_{p-1} $ is the $(p-1) \times (p-1)$ identity matrix.

\subsection{Evaluating Weights and the Likelihood Function}

The likelihood depends only on the observed data
\begin{equation}
f(\vecx_i) = \frac{1}{(2\pi)^{p/2} |\mSigma_{xx}|^{1/2} } \exp\left\{ -\frac{1}{2} (\vecx_i-\vecmu_x)' \mSigma_{xx}^{-1} (\vecx_i-\vecmu_x) \right\}.
\end{equation}
However, from Equation (\ref{quadratic vector result}) we have
\begin{equation} 
 \left( \hat{\vecy}_i - \vecmu\right)' \mSigma^{-1} \left( \hat{\vecy}_i - \vecmu\right) 
 = (\hat{\vecz}_i-\vecmu_{z.x})'\mSigma_{z.x}^{-1}(\hat{\vecz}_i-\vecmu_{z.x})
  + ( \vecx_i-\vecmu_{x})'\mSigma_{xx}^{-1}(\vecx_i-\vecmu_{x}). 
\end{equation}
Therefore,
\begin{equation}
(\vecx_i-\vecmu_x)' \mSigma_{xx}^{-1} (\vecx_i-\vecmu_x) 
\le
\left( \hat{\vecy}_i - \vecmu\right)' \mSigma^{-1} \left( \hat{\vecy}_i - \vecmu\right)
= 
\left( \hat{\vecy}_i - \vecmu\right)' \mXi \left( \hat{\vecy}_i - \vecmu\right)
\end{equation}
and we have equality when $\hat{\vecz}_i=\vecmu_{z.x}$. Our algorithm for $\hat{\vecz}_i$ will converge to $\vecmu_{z.x}$, so if we use this approximation for the likelihood and weights calculations they will also converge to the true quantities. 

To calculate $|\mSigma_{xx}|$, we use the relationship between Schur complements and the determinant, and the relationship between the inverse matrix and the Schur complement;
\begin{equation*} \label{{Schur  determent relation}}
\ln|\mSigma| =  \ln|\mSigma_{xx}| + \ln|\mSigma_{z.x}|  = \ln|\mSigma_{xx}| -\ln|\mXi_{zz}|  .
\end{equation*}
Alternatively, we could use our current estimate of $\mSigma_{z.x_i}$, namely, $\vecZ_{i}$. However, note that if the dimension of the missing data is larger than the dimension of the observed data for observation $i$ then it will better to calculate $|\mSigma_{xx}|$ directly.

\subsection{Model Selection}

The Bayesian information criterion \citep[BIC;][]{schwarz78} is used to select the number of components $G$ and the number of latent factors $q$. For a model with parameters $\vectheta$, $\text{BIC}=2l(\vecx,\hat\vectheta)-m\log n$,
where $l(\vecx,\hat\vectheta)$ is the maximized log-likelihood,
$\hat\vectheta$ is the maximum likelihood estimate of $\vectheta$, $m$
is the number of free parameters in the model, and $n$ is the
number of observations. The use of the BIC in mixture model selection was originally \citep{dasgupta98} based on an approximation to Bayes factors \citep{kass95}. The effectiveness of the BIC for choosing the number of factors in a factor analysis model has been established by \cite{lopes04}. 

\section{Analysis of the White Bread Data}\label{sec:white}
\subsection{The White Bread Data}
A total of $n=369$ consumers tasted six out of $12$ white breads in a BIB design. Taste was evaluated on the hedonic scale, so values in $\{1,2,\ldots,9\}$ are assigned to each tasted bread. 
For illustration, the first few rows of the data are shown in Table~\ref{tab:data}, where the bread brands are denoted $\text{A},\text{B},\ldots,\text{L}$. We fitted our mixture of factor analyzers model, with common factors, to these data using the PEM algorithm introduced herein. These models were fitted for $G=1,\ldots,6$ and $q=1,\ldots,3$, using multiple restarts.
\begin{table}[!ht]
\caption{\label{tab:data}The first six rows of the white bread data, where each consumer evaluates six breads using the hedonic scale.} 
\centering
\begin{tabular*}{1.0\textwidth}{@{\extracolsep{\fill}}l|cccccccccccc}
\hline
Consumer& A& B&  C&    D&    E&    F&    G&    H&    I&     J&     K&     L \\
\hline
1&    9&   &    8&    6&     &     &     &    9&     &      &     4&     8 \\
2&    3&   &    8&     &    7&     &    8&    7&    8&      &      &       \\
3&     &  8&    6&    7&     &     &     &     &    6&     9&     7&       \\
4&     &   &    5&    4&     &    6&     &    4&    3&     6&      &       \\
5&     &   &    7&    7&     &     &    8&    7&    6&      &     8&       \\
6&     &   &     &    8&     &     &    3&    4&    8&      &     7&     7 \\
\hline
\end{tabular*}
\end{table}

\subsection{Results}
\begin{figure}[!h]
\centering\includegraphics[width=6.75in]{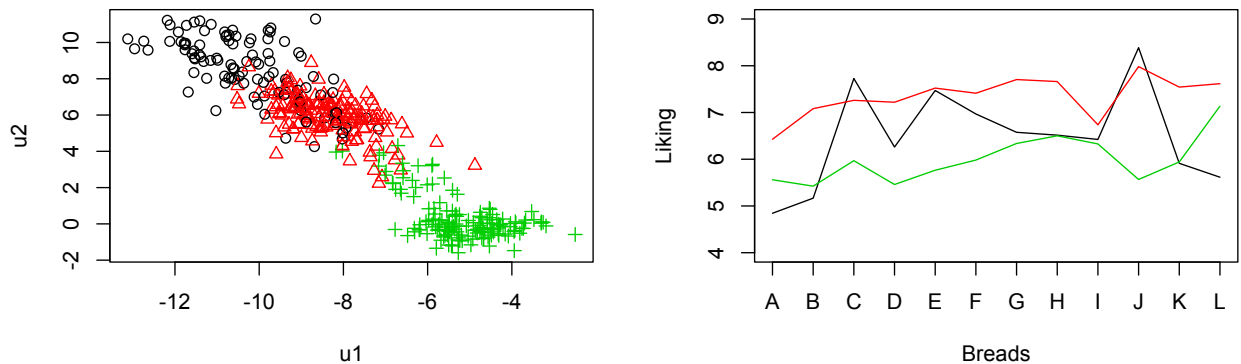} 
\caption{Plot of the two latent factors for the selected model, coloured by component (left), and a plot of the average liking scores for each of the breads separated by component (right).}
\label{bread3_solution}
\end{figure}

The results (Table~\ref{bic values}) show that the BIC selected a model with $G=3$ components and $q=2$ factors. Note that we also ran standard EM algorithms on these data and can confirm that they converged to the same results as our PEM algorithms. A plot of the two latent factors (Figure~\ref{bread3_solution}) shows the three components in the latent space. Because the classifications are based on maximum \textit{a posteriori} (MAP) probabilities, it is straightforward to provide the client with probabilities rather than hard group memberships; this might be particularly desirable for consumers near the cluster boundaries.
\begin{table}[!h]
\caption{\label{bic values} BIC values from our analysis of the white bread data, for $G=1,\ldots,6$ components and $q=1,\ldots,3$ latent factors.}
\centering{\begin{tabular*}{1.0\textwidth}{@{\extracolsep{\fill}}lrrr}
\hline
& \multicolumn{3}{c}{Number of Latent Factors}  \\
\cline{2-4}
$G$ & 1 & 2 & 3    \\
\hline
1& 5273.9& 5318.0& 5369.9 \\
2& 5176.1& 5136.0& 5193.1 \\ 
3& 5148.1& \textbf{5125.5}& 5244.1 \\ 
4& 5182.2& 5171.5& 5285.0 \\ 
5& 5223.1& 5288.1& 5341.7 \\ 
6& 5374.1& 5439.1& 5492.7 \\ 
\hline
\end{tabular*}}
\end{table}

In the same figure, there is also a plot of the mean liking scores for each bread for each of the three components. The red and green components seem to represent higher and lower scorers, respectively, with consumers within the black component exhibiting more variability in liking.  

Some interesting points emerge from inspection of the results. Notably, bread~J emerges as polarizing: it is strongly liked in the red and black groups and disliked in the green group. Interestingly, bread~J is the only ciabatta-style bread in the study and so it makes sense that its sensory properties will result in a relatively extreme liking response. This liking contrast is useful in differentiating groups of consumers because the objective of this research is to understand the sensory-based choice behaviour of consumers to define an optimum product for each liking cluster.
Bread~I is also interesting, in that it is the one bread for which consumers in all three groups seem to converge to the similar liking scores. Bread~I is the sweetest, most flavourful bread in the study; it is also firm, dense, moist, and chewy. This is an unusual combination of characteristics and one would expect it to stand out. The fact that it stood out by not differentiating consumers in this study is itself interesting in the process of trying to understand the sensory-based choice behaviour of consumers.

\subsection{Comparing PEM and EM}\label{sec:sim}
\begin{figure}[!ht] 
\centering\includegraphics[width=6.5in]{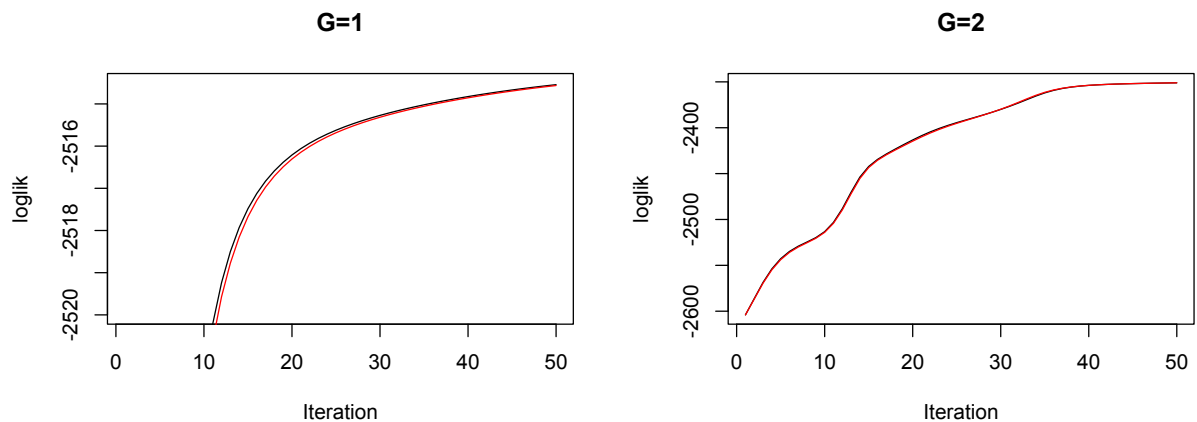} 
\caption{Plot of the log-likelihood for $G=1$ and $G=2$ for PEM (red line) and EM (black line) algorithms.}
\label{compare_loglik}
\end{figure}

The analysis of the bread data was repeated using a standard EM algorithm for parameter estimation. The results were the same, as we would expect. Figure~\ref{compare_loglik} illustrates the progression of the EM and PEM algorithms with $G=1$ and $G=2$ components, respectively, and $q=2$ latent factors. As expected, both algorithms converge to the same solution in an almost identical fashion. 

\section{Discussion}\label{sec:disc}

We developed an approach for clustering incomplete BIB data from consumer tasting of $12$ different commercial white breads. Our clustering approach is based on a parsimonious mixture of factor analyzers model, where the factor loading matrices are constrained to be equal across groups. The problem of missing data is handled along with parameter estimation within an partial EM algorithm framework. Rather than simple imputation, this PEM algorithm approach  effectively imputes missing data at each iteration based on current component membership probabilities; this is a natural approach as missing values are filled in based on complete values in observations that are in some sense similar (i.e., in the same component). Our PEM algorithm is much more computationally efficient than the standard EM algorithm for this, and any such, missing data problem. The PEM is shown to retain the monotonicity property and, thus, retains the same convergence properties as the EM algorithm.
Three benefits are achieved through this approach: the quality of data that are collected prior to fatigue is improved; the method of substituting missing data reflects the sensory preferences of each consumer, which permits robust cluster assignment; and the collection of incomplete-block data reduces the cost, time, and materials required for this type of study. 

We introduce a new variation of the EM algorithm called the PEM algorithm. The many varieties of the EM mainly focus on the M-step: the expectation-conditional maximization (ECM) algorithm \citep{meng93}, the ECM either (ECME) algorithm \citep{liu94}, the alternating ECM (AECM) algorithm, and others. Few examine different ways of partially updating the E-step. \cite{neal1998} give several possible methods to update the missing sufficient statistics. All of the methods suggest something along the lines of fully updating a partial set of the  missing sufficient statistics. When the E-step is intractable, \cite{wei90} suggest approximating the E-step by simulating $m$ observations from the conditional distribution of the missing data given the observed data. This version of EM algorithm is called Monte Carlo EM (MCEM). Prior to MCEM, \cite{celeux94} suggested using stochastic EM (SEM), which is the same as MCEM with $m=1$. Other variations on approximating the E-step have been introduced, such as MCEM using rejection sampling, importance sampling, and Markov chain Monte Carlo. 
The PEM algorithm presented here is similar to ECM in which we have `conditional' E-steps instead of conditional M-steps. These conditional E-steps are computationally cheaper than using a complete or full E-step. 






\section*{Acknowledgements}
This work was supported by a grant-in-aid from Compusense Inc.\ and by a Collaborative Research and Development grant from the Natural Sciences and Engineering Research Council of Canada. 

\appendix
\section{Some Mathematical Details}
\subsection{Schur Complement Relation in Quadratic Form } \label{vector quadratic form}
Suppose we have a positive-definite symmetric matrix $\vecS$ and a vector $\vecy$ with decompositions 
\begin{equation*}
\vecy=\left[\begin{array}{c} \vecy_1 \\ \vecy_2 \end{array}\right]\qquad\text{and}\qquad 
\vecS = \left[ \begin{array}{cc} \vecS_{11} &  \vecS_{12} \\  \vecS_{21}  &  \vecS_{22} \end{array}\right],
\end{equation*}
then
\begin{equation*}
\vecy'\vecS^{-1}\vecy = 
\vecy_1'\vecS_{11}^{-1}\vecy_1 
+ (\vecy_2 - \vecS_{21}\vecS_{11}^{-1}\vecy_1)' \vecS_{22.1}^{-1} (\vecy_2 - \vecS_{21}\vecS_{11}^{-1}\vecy_1),
\end{equation*}
where $\vecS_{22.1} =\vecS_{22} - \vecS_{21} \vecS_{11}^{-1}\vecS_{12} $. 

\subsection{A Matrix Minimization Problem} \label{matrix quadratic form}
Suppose we have a positive-definite symmetric matrix $\vecS$ with decomposition 
\begin{equation*}
\vecS =  \left[\begin{array}{cc} \vecS_{11} & \vecS_{12} \\ \vecS_{21} & \vecS_{22}\end{array}\right].
\end{equation*}
We then have the following property for a function $\gamma$,
\begin{equation} \label{matrix minimization}
h(\mTheta_{11} ) = \mbox{tr} \left\{\left( \vecS_{11} - \mTheta_{11}, \vecS_{12} \right)  \vecS^{-1}  \left( \vecS_{11} - \mTheta_{11}, \vecS_{12} \right) \right\}  
\;\;
\ge
\;\;
\mbox{tr} \left\{\vecS_{22}^{-1} \vecS_{12} \vecS_{12}'\right\}.
\end{equation}
Equality holds when $\mTheta_{11} = \vecS_{11} - \vecS_{12} \vecS_{12} \vecS_{21} $. Therefore, $h(\mTheta_{11} )$ is minimized by the Schur complement  $\mTheta_{11} = \vecS_{11} - \vecS_{12} \vecS_{12} \vecS_{21} $. Now, if we define
\begin{equation}
\mTheta_ =  \left[\begin{array}{cc} \mTheta_{11} & 0 \\ 0 & 0 \end{array}\right],
\end{equation}
then
\begin{equation}
h(\mTheta_{11}) = \mbox{tr}\left\{\left( \vecS - \mTheta \right)  \vecS^{-1}  \left( \vecS - \mTheta \right) \right\}  =  \gamma(\mTheta_{11} ) + \mbox{tr} \left\{\vecS_{22}^{-1} \vecS_{22} \vecS_{22} \right\}.
\end{equation}
Because the right-hand term does not depend on $\mTheta_{11}$, $h(\mTheta_{11} )$ has the same minimum as $g(\mTheta_{11} )$; i.e., the Schur complement $\mTheta_{11} = \vecS_{11} - \vecS_{12} \vecS_{12} \vecS_{21} $. Therefore, a minimization algorithm based on the function $h$ is equivalent to minimizing $\gamma$. We minimize $h$ using a conditional minimization algorithm (by column/row) based on Equation~(\ref{matrix minimization}). 

\bibliographystyle{chicago}
\bibliography{mybib}

 \end{document}